\documentclass[aps,prl,twocolumn,showpacs,groupedaddress,amsmath,amssymb,floatfix]{revtex4-1}
\usepackage{graphicx}

\begin{document}


\title{Time resolved viscoelastic properties during structural arrest and aging of a colloidal glass}


\author{Ajay Singh Negi}
\email[]{ajay.negi@yale.edu}
\affiliation{Department of Chemical Engineering, Yale University,
New Haven CT 06511}
\author{Chinedum O. Osuji}
\email[]{chinedum.osuji@yale.edu}
\affiliation{Department of Chemical Engineering, Yale University,
New Haven CT 06511}
\affiliation{}


\date{\today}


\begin{abstract}
Evolution of the energy landscape during physical aging of glassy
materials can be understood from the frequency and strain dependence
of the shear modulus but the non-stationary nature of these systems
frustrates investigation of their instantaneous underlying properties. Using a series
of time dependent measurements we systematically reconstruct the
frequency and strain dependence as a function of age for a repulsive
colloidal glass undergoing structural arrest. In this manner, we are
able to unambiguously observe the structural relaxation time, which
increases exponentially with sample age at short times. The yield
stress varies logarithmically with time in the arrested state,
consistent with recent simulation results, whereas the yield strain
is nearly constant in this regime. Strikingly, the frequency
dependence at fixed times can be rescaled onto a master curve,
implying a simple connection between the aging of the system and the
change in the frequency dependent modulus.
\end{abstract}

\maketitle

Time evolution of the dynamic and mechanical properties is a
hallmark of many disordered materials. Such physical aging is found
in polymeric and molecular glass formers below the glass transition
temperature, $T_g$, as well as in dense colloidal systems at rest
\cite{Angell_McKenna_review_2000,Sciortino_review_2005}. The
overarching commonality is the structural arrest that occurs at the
glass transition which is manifest in the dramatic slowing of the
system's dynamics. The boundary of the glassy state in colloidal
suspensions is defined by volume fraction, $\varphi$, whereas in
other structural glasses the boundary is thermal and defined by a
nominal $T_g$. In such systems, structural arrest and aging occur
following a rapid temperature quench from the equilibrium liquid
state, charting the slow evolution of the glass towards a presumed
long-time equilibrium or time-invariant state. In colloidal systems
however, with the exception of recent thermoresponsive materials,
for fixed $\varphi$, the entry into the glassy state is mechanically
controlled by the application and subsequent removal of shear
stresses well in excess of the material's yield stress. This has
variously been called shear melting and rejuvenation wherein the
effective thermalizing role of stress is utilized to produce a well
defined initial state for the colloidal glass
\cite{Cloitre_PRL2000,Joshi_2009modeling,Vlassopoulos_aging2009}.
Rapid cessation of flow produces in effect a mechanical quench from
a free-flowing fast relaxing liquid to a metastable fluid which
quickly becomes arrested as it enters the glassy or jammed state and
undergoes subsequent aging, as shown by Ovarlez and Coussot
\cite{Ovarlez_Coussot_PRE2007}.

Progress has been made in simulations of thermally quenched LJ
\cite{Kob_Barrat_PRL1997,Varnik_Barrat2004,Rottler_PRL2005} and hard
sphere glasses \cite{Puertas_aging2010}, as well as in the
phenomenological soft glassy rheology (SGR) model
\cite{Fielding_Cates_2000} which utilizes an effective temperature
concept. Comparatively less is known regarding the evolution of
colloidal glasses at short times following a mechanical quench as
the system undergoes structural arrest from a metastable fluid.
Access to the time evolution of the frequency and strain dependent
complex shear modulus, $G^*(\omega,t)$ and $G^*(\gamma,t)$, during
structural arrest would provide valuable insight into the nature of
the colloidal glass transition, particularly as these are properties
of interest for theoretical frameworks, including SGR. Recent
studies have succeeded in characterizing $G(t_w,\omega)$ in freshly
mixed suspensions at times greater than 30 mins
\cite{Jabbari-Farouji_PRE2008}. Measurement of these properties at
shorter times, however, is a significant challenge in these fast
aging materials, and such time resolved properties have not been
reported to date. In this Letter we consider the time dependence of
$G^*(\omega)$ and $G^*(\gamma)$ for a model colloidal system with
repulsive interactions. A systematic series of time dependent
measurements facilitates reconstruction of the frequency and strain
dependence as a function of age after cessation of a strong
rejuvenating flow. We are able to observe the structural relaxation
time for systems of small age and find that it scales exponentially
with elapsed time. The yield stress increases logarithmically with
age while viscous dissipation during yielding increases linearly.
Remarkably, $G'(\omega)$ and $G''(\omega)$ at fixed times can be
shifted onto a master curve that describes the frequency dependence
of the system over all times considered.

Our system is a repulsive colloidal glass of charged disc-like clay
particles, Laponite, in water. At low ionic strength, long-range
Coulombic repulsion produces arrested states at low concentrations
of a few wt.\% which display aging
\cite{Willenbacher1996,bonn_epl_1998}. Laponite XLG (Southern Clay
Products) was dried overnight and then mixed with water at pH 9.5 to
produce a 3.5 wt.\% suspension. Suspensions were mixed by vortexing
for 2 minutes followed by sonication for 20 minutes and then allowed
to stand quiescently for 4 days. During this time, the clay
particles become fully hydrated and form a clear gel-like material.
Studies were performed on this resultant material and thus reflect
dynamics of the arrested glass and not any changes due to initial
hydration \cite{Joshi_cage_formation_2007}. Measurements were
conducted on an Anton-Paar MCR301 rheometer in strain-controlled
mode in a cone-plate geometry (1$^{\circ}$, 50 mm). Samples were
subjected to a strong rejuvenating flow at $\dot\gamma=3000s ^{-1}$
for 100 s and then mechanically quenched by a rapid linear cessation
of flow over 1 s. This resulted in a reproducible and well defined
initial state. A series of time sweeps were conducted either at a
fixed strain of $\gamma=1$\% for the frequency dependent experiments
for 1000 s, or at a fixed frequency of $\omega=10$ rad/s. for strain
dependent studies for 600 s. At the end of each time sweep samples
were subjected to the rejuvenating flow step before measurements at
the next frequency or strain. A thin film of mineral oil was
successfully used at the edge of the cone to suppress evaporation.

\begin{figure}[h]
\includegraphics[width=75mm,scale=1]{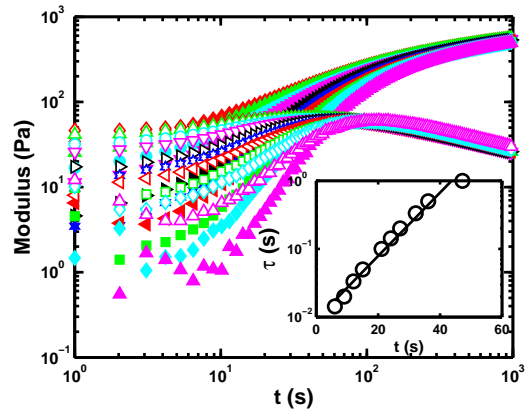}
\caption{Evolution of storage (filled symbols) and loss (open
symbols) moduli at different frequencies, from 70 rad/s
(largest $G'$) to 1 rad/s (smallest $G'$). Inset: Relaxation
time as a function of age. The age is $t_c$, the cross over
time and $\tau=1/\omega$, the inverse of the sampling
frequency. The line is a fit to an exponential function
with a characteristic time $\approx 10$s.
\label{moduli_evolution_frequency}}
\end{figure}

The strong shear flow in the rejuvenating step produces a liquid
state from an initially solid-like material. At short times after
flow cessation, the system exhibits liquid-like behavior, with the
loss modulus exceeding the storage modulus, $G''>G'$. Both moduli
increase rapidly with time, but the rate of increase of $G'$ is
faster such there is a crossover at a characteristic time, $t_c$,
during the experiment, Fig. \ref{moduli_evolution_frequency}. This
time ranged from about 5 to 50 seconds, with the shortest times
encountered for the highest frequencies. In all cases, $t_c\gg
1/\omega$, ensuring that data are not affected by sample evolution
during a single strain cycle. At long times $G^*\sim\ln(t)$ but the
distinction from a weak power-law dependence where $G'\sim t^{0.5}$
and $G''\sim t^{-0.4}$ is small.

\begin{figure}[h]
\includegraphics[width= 75mm, scale=1]{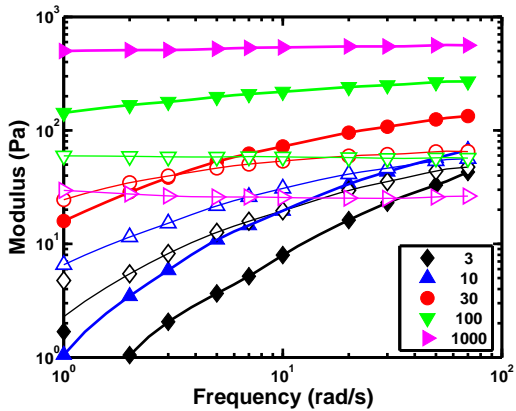}
\caption{Reconstructed frequency dependence of the storage and loss
moduli, at 5 times shown in seconds. Lines are visual
guides. \label{frequency_dependent_moduli}}
\end{figure}

The crossover time is the characteristic timescale for the
structural arrest of the system following the perturbation produced
by the rejuvenating shear flow \cite{Osuji_Negi_arrest_EPL2010}.
Microscopically, it is related to the timescale for the formation of
particle cages responsible for dynamic arrest
\cite{Sciortino_review_2005}. This timescale is not the same as the
structural relaxation timescale of the system. The structural
relaxation time, $\tau$, is given by the inverse of the sampling
frequency when the sample is at an age equal to the structural
arrest time, $t_c$, inset Fig. \ref{moduli_evolution_frequency}.
This point is made clear in the frequency representation of the
data. The frequency dependence at fixed sample age is reconstructed
from the time-dependent data across different frequencies. For
clarity, 5 times are selected, Fig.
\ref{frequency_dependent_moduli}. At short times, $t\lesssim 10 s$,
the system displays a frequency dependence that is typical of a
liquid-like state, with $G''>G'$.  At intermediate times, $10<t<100
s$, the frequency dependence lessens and there are crossovers
between $G'$ and $G''$ that falls within the accessible frequency
space. Finally, at long times, $t>100 s$ the frequency dependence
has nearly vanished as the relaxation frequency for the system is
far outside the observation window, and the system is well arrested.
A plot of the inverse of the cross-over frequencies as a function of
sample age simply replicates the inset plot of Fig.
\ref{moduli_evolution_frequency}. The relaxation time varies
exponentially with sample age, $\tau\sim\exp(t/\tau_a)$, with a
characteristic time $\tau_a\approx 10s$. Exponential evolution of
relaxation time has been observed in Laponite suspensions during the
liquid-solid transition following the initial mixing of clay
aggregates with water \cite{Bonn_PRE2001,COO:Bellour_Munch}. In
those cases the transition occurred on timescales of several hours
and partially reflects the initial hydration of the system. In the
present case, the transition occurs on timescales less than $10^2$ s
in an already well hydrated system. The information contained in the
extracted frequency dependent spectra is quite compelling. The
spectra represent the instantaneous underlying frequency response of
the system as encountered during the rapid aging from a liquid-like
to solid-like state.

\begin{figure}[h]
\includegraphics[width= 75mm, scale=1]{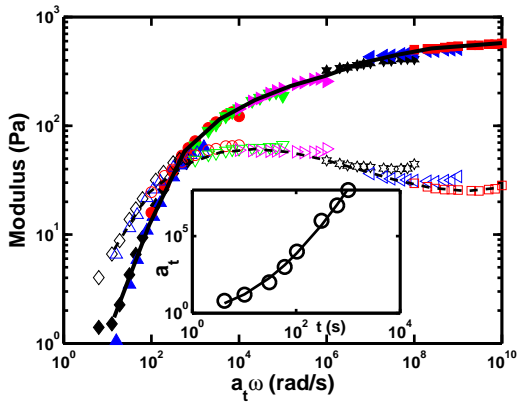}
\caption{Master curve of storage and loss moduli (filled and open
symbols) as a function of reduced frequency. Lines are shown as
visual guides. Inset: Shift factors $a_t$ as a function of system
age. The line  is a fit to $a_t\sim -a\exp(-t/\tau_a^s)+bt^{x}+c$
with $\tau_a^s\approx 9$s and $x=1.8$.
\label{master_curve_frequency}}
\end{figure}

Remarkably, frequency dependent data at fixed times can be shifted
to construct a master curve. Data at short times as expected are
shifted to lower frequencies where the response of the system is
more liquid-like, and conversely long-time data are shifted to
higher frequencies as well aged systems display solid-like behavior.
The master curve is constructed using only horizontal shifts, Fig.
\ref{master_curve_frequency}, along with shift factors inset. At
short times the shift factor dependence on age mirrors that of the
relaxation timescale, with an exponential form. This is followed by
a power-law regime at long times. The shift factors are fit as
$a_t\sim -a\exp(-t/\tau_a^s)+bt^{x}+c$. The characteristic time for
the exponential component here is $\tau_a^s\approx 9s$ and $x\approx
1.8$. The same shift factors have been used to scale both the
storage and loss moduli. The correspondence between the
characteristic times, $\tau_a^s\approx\tau_a$, suggests strongly
that the evolving structural relaxation underlies the rescaling of
the data at short times.

\begin{figure}[h]
\includegraphics[width= 75mm, scale=1]{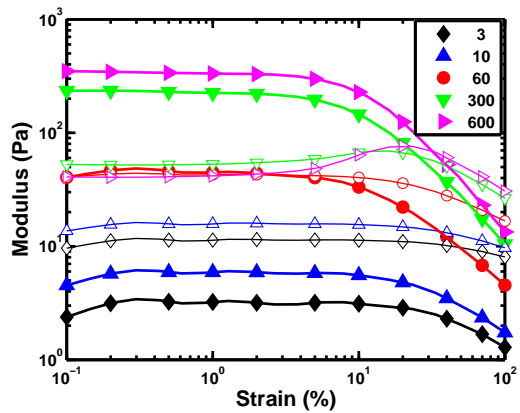}
\caption{Reconstructed strain dependent storage and loss moduli, at
different system ages in seconds as indicated. Lines are shown as visual guides.
\label{strain_dependent_moduli}}
\end{figure}

We interpret our data as indicative of two distinct processes. The
first is structural arrest and the second is aging. Structural
arrest as discussed is associated particle localization and the
process mediates the transition from the metastable fluid to the
arrested glass. Beyond this caging regime, the system settles into
power-law aging wherein the evolution of properties occurs via the
small scale cage rearrangements that are presumed to characterize
the aging of structural glasses. This regime is characterized by the
system's exploration of its energetic landscape and continually
slowing evolution towards a stationary state. These results are
qualitatively similar to those of Bellour {\emph et al.} where a
transition from exponential to power-law aging of the relaxation
time was observed in freshly prepared suspensions after several
hours \cite{COO:Bellour_Munch}. The strain dependence of the system
modulus similarly displays two regimes. The same procedure is
applied to time sweep data conducted at different strains to extract
the underlying strain dependence of the system at fixed times. Again
for clarity, a small subset of such reconstructed data sets is
shown, Fig. \ref{strain_dependent_moduli}.

\begin{figure}[h]
\includegraphics[width= 75mm,
scale=1]{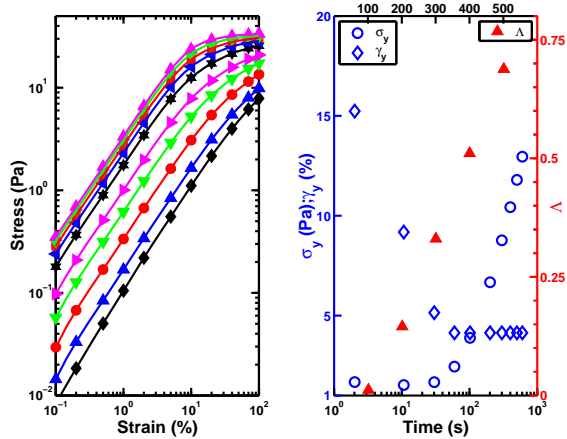}
\caption{Left: Stress versus strain at different system ages of 1
(bottom), 10, 30, 60, 100, 200, 300, 400, 500 and 600 (top) seconds.
Lines are shown as visual guides. Right: Semi-log plot of $\sigma_y$
and $\gamma_y$ as a function of system age on left axis. On right
axis: time dependent plot of normalized dissipation
$\Lambda=G''_{peak}/\langle G''_{linear}\rangle-1$ during yielding.
\label{stress_strain_yield}}
\end{figure}

At short times, the system is a fluid with $G''>G'$ whereas at long
times it displays the strain sweep typical of a soft glassy solid,
with $G'>G''$ and a peak in the loss modulus during yielding.
Examination in terms of stress-strain curves permits extraction of
the yield point defined for the current purpose as the strain at
which the slope (the shear modulus) deviates by more than
$\epsilon_{dev}=5\%$, Fig. \ref{stress_strain_yield}. The data are
robust against variation of $\epsilon_{dev}$ from 2.5\% to 10\%. The
yield stress and yield strain, $\sigma_y$ and $\gamma_y$, show
opposing tendencies. At short times, $\sigma_y$ is quite independent
of system age, but shows a marked upturn around $t=20-30$s after
which it evolves logarithmically with time. By contrast, $\gamma_y$
continually decreases with time, before an asymptotic approach near
$t=100$s to a relatively constant value, $\gamma_y\approx 4\%$,
which is maintained at long times. Logarithmic increase of yield
stress with age is known from experiments on polymer glasses
\cite{Chow1993stress} and has been recovered in simulations on
binary LJ glasses \cite{Varnik_Barrat2004,Rottler_PRL2005} and
polymers \cite{Schweizer_aging_polymers_PRE2008}. As noted in
\cite{Rottler_PRL2005}, the short time domain where $\sigma_y$ is
independent of age is not a common or easily accessed situation
experimentally. The data of Fig. \ref{stress_strain_yield} however
reproduce this behavior quite well and provide experimental
verification of the simulation findings. The results are coherent
with the framework developed from the frequency dependent data. They
corroborate a scenario in which the system evolves first via
structural arrest due to cage formation, and then later via aging
which continues to increase the modulus and yield stress of the
glass. The yielding of structural glasses is known to occur via a
peak in the loss modulus, which is associated with viscous
dissipation during the cage breaking process
\cite{Miyazaki_Wyss2006}. Prior work has established that the size
of this peak, as well as the stress overshoot in start-up flow is
correlated to the time at rest or age of the system
\cite{Fielding_Cates_2000,Derec_PRE2003}. Here, we resolve the
evolution of this viscous dissipation with time, including at short
absolute times, as well as times that are short relative to the
timescale for conducting a traditional strain sweep measurement. We
define a normalized quantity, $\Lambda=G''_{peak}/\langle
G''_{linear}\rangle-1$ based on the ratio of the loss modulus peak
to its average value in the linear regime. The data reveal an
absence of measurable dissipation at short times, $t< 100$s, and a
linear increase in $\Lambda$ with time thereafter.

These experiments have characterized in detail the frequency and
strain dependence of a glassy system during structural arrest and
subsequent aging. They rely on the fidelity of the strong
rejuvenating flow in consistently returning the system to a well
defined fluid-like initial state from which it subsequently evolves.
This is in contrast to studies in which relatively weak rejuvenating
flows were used, resulting in near-linear \cite{Hebraud_PRE2005} or
slightly super-linear
\cite{Viasnoff_PRL2002,Ozon_PRE2003,Ruocco_PRE2007} growth of the
relaxation time with age. The use of much higher flow rates to
rejuvenate the sample here resulted in access to a much more fluid
state than considered in prior reports where strains and shear rates
did not exceed $\gamma\approx 300\%$ and $\dot\gamma\approx 150
s^{-1}$. This may account for the faster rate of aging following
structural arrest and underscores the role of flow history as a
determinant of the aging response \cite{Osuji_Negi_JRheol2010}.

The present data provide significant insight regarding the dynamics
of soft glassy materials in regimes of interest that have been
relatively unexplored. They reveal a rapid transition from the
arrest regime to an aging regime with a typical timescale that is
reflected in the frequency and strain dependent properties. The
frequency data permit the unambiguous resolution of the structural
relaxation time as a function of age, and display a universal
scaling that is consistent with exponential and power-law
dependencies in the arrest and aging regimes, respectively. The
strain dependent moduli exhibit the same liquid to solid-like
transition, with a decreasing yield strain and logarithmically
increasing yield stress as a function of age, with marked changes in
the yield stress near the timescale associated with structural
arrest, $\tau_a$. Concurrently, enhanced dissipation on yielding is
seen to be a feature only of the arrested state, with the magnitude
of this dissipation increasing linearly with age. To the best of our
knowledge, these are the first experimental data that chart the
dynamics and yielding behavior of a colloidal system undergoing
structural arrest and subsequent aging at short times where many key
features emerge. These results frame a useful picture of the
dynamics of an out of equilibrium system in the transition from a
metastable liquid to an aging glass.


\acknowledgements{The authors are grateful to K. Schweizer, G.
Petekidis and D. Vlassopoulos for insightful discussions, and
acknowledge funding by NSF under CBET-0828905}

\bibliography{frequency_strain_dependence_aging}

\begin{thebibliography}{27}%
\makeatletter
\providecommand \@ifxundefined [1]{%
 \@ifx{#1\undefined}
}%
\providecommand \@ifnum [1]{%
 \ifnum #1\expandafter \@firstoftwo
 \else \expandafter \@secondoftwo
 \fi
}%
\providecommand \@ifx [1]{%
 \ifx #1\expandafter \@firstoftwo
 \else \expandafter \@secondoftwo
 \fi
}%
\providecommand \natexlab [1]{#1}%
\providecommand \enquote  [1]{``#1''}%
\providecommand \bibnamefont  [1]{#1}%
\providecommand \bibfnamefont [1]{#1}%
\providecommand \citenamefont [1]{#1}%
\providecommand \href@noop [0]{\@secondoftwo}%
\providecommand \href [0]{\begingroup \@sanitize@url \@href}%
\providecommand \@href[1]{\@@startlink{#1}\@@href}%
\providecommand \@@href[1]{\endgroup#1\@@endlink}%
\providecommand \@sanitize@url [0]{\catcode `\\12\catcode `\$12\catcode
  `\&12\catcode `\#12\catcode `\^12\catcode `\_12\catcode `\%12\relax}%
\providecommand \@@startlink[1]{}%
\providecommand \@@endlink[0]{}%
\providecommand \url  [0]{\begingroup\@sanitize@url \@url }%
\providecommand \@url [1]{\endgroup\@href {#1}{\urlprefix }}%
\providecommand \urlprefix  [0]{URL }%
\providecommand \Eprint [0]{\href }%
\@ifxundefined \urlstyle {%
  \providecommand \doi  [0]{\begingroup \@sanitize@url \@doi}%
  \providecommand \@doi [1]{\endgroup \@@startlink {\doibase
  #1}doi:\discretionary {}{}{}#1\@@endlink }%
}{%
  \providecommand \doi  [0]{doi:\discretionary{}{}{}\begingroup
  \urlstyle{rm}\Url }%
}%
\providecommand \doibase [0]{http://dx.doi.org/}%
\providecommand \Doi [0]{\begingroup \@sanitize@url \@Doi }%
\providecommand \@Doi  [1]{\endgroup\@@startlink{\doibase#1}\@@Doi}%
\providecommand \@@Doi [1]{#1\@@endlink}%
\providecommand \selectlanguage [0]{\@gobble}%
\providecommand \bibinfo  [0]{\@secondoftwo}%
\providecommand \bibfield  [0]{\@secondoftwo}%
\providecommand \translation [1]{[#1]}%
\providecommand \BibitemOpen [0]{}%
\providecommand \bibitemStop [0]{}%
\providecommand \bibitemNoStop [0]{.\EOS\space}%
\providecommand \EOS [0]{\spacefactor3000\relax}%
\providecommand \BibitemShut  [1]{\csname bibitem#1\endcsname}%
\bibitem [{\citenamefont {Angell}\ \emph {et~al.}(2000)\citenamefont {Angell},
  \citenamefont {Ngai}, \citenamefont {McKenna}, \citenamefont {McMillan},\
  and\ \citenamefont {Martin}}]{Angell_McKenna_review_2000}%
  \BibitemOpen
  \bibfield  {author} {\bibinfo {author} {\bibfnamefont {C.}~\bibnamefont
  {Angell}}, \bibinfo {author} {\bibfnamefont {K.}~\bibnamefont {Ngai}},
  \bibinfo {author} {\bibfnamefont {G.}~\bibnamefont {McKenna}}, \bibinfo
  {author} {\bibfnamefont {P.}~\bibnamefont {McMillan}}, \ and\ \bibinfo
  {author} {\bibfnamefont {S.}~\bibnamefont {Martin}},\ }\href@noop {}
  {\bibfield  {journal} {\bibinfo  {journal} {J. Appl. Phys.},\ }\textbf
  {\bibinfo {volume} {88}},\ \bibinfo {pages} {3113} (\bibinfo {year}
  {2000})}\BibitemShut {NoStop}%
\bibitem [{\citenamefont {Sciortino}\ and\ \citenamefont
  {Tartaglia}(2005)}]{Sciortino_review_2005}%
  \BibitemOpen
  \bibfield  {author} {\bibinfo {author} {\bibfnamefont {F.}~\bibnamefont
  {Sciortino}}\ and\ \bibinfo {author} {\bibfnamefont {P.}~\bibnamefont
  {Tartaglia}},\ }\href@noop {} {\bibfield  {journal} {\bibinfo  {journal}
  {Adv. Phys.},\ }\textbf {\bibinfo {volume} {54}},\ \bibinfo {pages} {471}
  (\bibinfo {year} {2005})}\BibitemShut {NoStop}%
\bibitem [{\citenamefont {Cloitre}\ \emph {et~al.}(2000)\citenamefont
  {Cloitre}, \citenamefont {Borrega},\ and\ \citenamefont
  {Leibler}}]{Cloitre_PRL2000}%
  \BibitemOpen
  \bibfield  {author} {\bibinfo {author} {\bibfnamefont {M.}~\bibnamefont
  {Cloitre}}, \bibinfo {author} {\bibfnamefont {R.}~\bibnamefont {Borrega}}, \
  and\ \bibinfo {author} {\bibfnamefont {L.}~\bibnamefont {Leibler}},\
  }\href@noop {} {\bibfield  {journal} {\bibinfo  {journal} {Phys. Rev.
  Lett.},\ }\textbf {\bibinfo {volume} {85}},\ \bibinfo {pages} {4819}
  (\bibinfo {year} {2000})}\BibitemShut {NoStop}%
\bibitem [{\citenamefont {Joshi}(2009)}]{Joshi_2009modeling}%
  \BibitemOpen
  \bibfield  {author} {\bibinfo {author} {\bibfnamefont {Y.}~\bibnamefont
  {Joshi}},\ }\href@noop {} {\bibfield  {journal} {\bibinfo  {journal}
  {Industrial \& Engineering Chemistry Research},\ }\textbf {\bibinfo {volume}
  {48}},\ \bibinfo {pages} {8232} (\bibinfo {year} {2009})}\BibitemShut
  {NoStop}%
\bibitem [{\citenamefont {Christopoulou}\ \emph {et~al.}(2009)\citenamefont
  {Christopoulou}, \citenamefont {Petekidis}, \citenamefont {Erwin},
  \citenamefont {Cloitre},\ and\ \citenamefont
  {Vlassopoulos}}]{Vlassopoulos_aging2009}%
  \BibitemOpen
  \bibfield  {author} {\bibinfo {author} {\bibfnamefont {C.}~\bibnamefont
  {Christopoulou}}, \bibinfo {author} {\bibfnamefont {G.}~\bibnamefont
  {Petekidis}}, \bibinfo {author} {\bibfnamefont {B.}~\bibnamefont {Erwin}},
  \bibinfo {author} {\bibfnamefont {M.}~\bibnamefont {Cloitre}}, \ and\
  \bibinfo {author} {\bibfnamefont {D.}~\bibnamefont {Vlassopoulos}},\ }\Doi
  {10.1098/rsta.2009.0166} {\bibfield  {journal} {\bibinfo  {journal} {Philos.
  Trans. R. Soc. London, Ser. A},\ }\textbf {\bibinfo {volume} {367}},\
  \bibinfo {pages} {5051} (\bibinfo {year} {2009})}\BibitemShut {NoStop}%
\bibitem [{\citenamefont {Ovarlez}\ and\ \citenamefont
  {Coussot}(2007)}]{Ovarlez_Coussot_PRE2007}%
  \BibitemOpen
  \bibfield  {author} {\bibinfo {author} {\bibfnamefont {G.}~\bibnamefont
  {Ovarlez}}\ and\ \bibinfo {author} {\bibfnamefont {P.}~\bibnamefont
  {Coussot}},\ }\Doi {10.1103/PhysRevE.76.011406} {\bibfield  {journal}
  {\bibinfo  {journal} {Phys. Rev. E},\ }\textbf {\bibinfo {volume} {76}},\
  \bibinfo {pages} {011406} (\bibinfo {year} {2007})}\BibitemShut {NoStop}%
\bibitem [{\citenamefont {Kob}\ and\ \citenamefont
  {Barrat}(1997)}]{Kob_Barrat_PRL1997}%
  \BibitemOpen
  \bibfield  {author} {\bibinfo {author} {\bibfnamefont {W.}~\bibnamefont
  {Kob}}\ and\ \bibinfo {author} {\bibfnamefont {J.-L.}\ \bibnamefont
  {Barrat}},\ }\Doi {10.1103/PhysRevLett.78.4581} {\bibfield  {journal}
  {\bibinfo  {journal} {Phys. Rev. Lett.},\ }\textbf {\bibinfo {volume} {78}},\
  \bibinfo {pages} {4581} (\bibinfo {year} {1997})}\BibitemShut {NoStop}%
\bibitem [{\citenamefont {Varnik}\ \emph {et~al.}(2004)\citenamefont {Varnik},
  \citenamefont {Bocquet},\ and\ \citenamefont {Barrat}}]{Varnik_Barrat2004}%
  \BibitemOpen
  \bibfield  {author} {\bibinfo {author} {\bibfnamefont {F.}~\bibnamefont
  {Varnik}}, \bibinfo {author} {\bibfnamefont {L.}~\bibnamefont {Bocquet}}, \
  and\ \bibinfo {author} {\bibfnamefont {J.-L.}\ \bibnamefont {Barrat}},\ }\Doi
  {10.1063/1.1636451} {\bibfield  {journal} {\bibinfo  {journal} {J. Chem.
  Phys.},\ }\textbf {\bibinfo {volume} {120}},\ \bibinfo {pages} {2788}
  (\bibinfo {year} {2004})}\BibitemShut {NoStop}%
\bibitem [{\citenamefont {Rottler}\ and\ \citenamefont
  {Robbins}(2005)}]{Rottler_PRL2005}%
  \BibitemOpen
  \bibfield  {author} {\bibinfo {author} {\bibfnamefont {J.}~\bibnamefont
  {Rottler}}\ and\ \bibinfo {author} {\bibfnamefont {M.~O.}\ \bibnamefont
  {Robbins}},\ }\Doi {10.1103/PhysRevLett.95.225504} {\bibfield  {journal}
  {\bibinfo  {journal} {Phys. Rev. Lett.},\ }\textbf {\bibinfo {volume} {95}},\
  \bibinfo {pages} {225504} (\bibinfo {year} {2005})}\BibitemShut {NoStop}%
\bibitem [{\citenamefont {Puertas}(2010)}]{Puertas_aging2010}%
  \BibitemOpen
  \bibfield  {author} {\bibinfo {author} {\bibfnamefont {A.}~\bibnamefont
  {Puertas}},\ }\href@noop {} {\bibfield  {journal} {\bibinfo  {journal} {J.
  Phys. Condens. Matter},\ }\textbf {\bibinfo {volume} {22}},\ \bibinfo {pages}
  {104121} (\bibinfo {year} {2010})}\BibitemShut {NoStop}%
\bibitem [{\citenamefont {Fielding}\ \emph {et~al.}(2000)\citenamefont
  {Fielding}, \citenamefont {Sollich},\ and\ \citenamefont
  {Cates}}]{Fielding_Cates_2000}%
  \BibitemOpen
  \bibfield  {author} {\bibinfo {author} {\bibfnamefont {S.~M.}\ \bibnamefont
  {Fielding}}, \bibinfo {author} {\bibfnamefont {P.}~\bibnamefont {Sollich}}, \
  and\ \bibinfo {author} {\bibfnamefont {M.~E.}\ \bibnamefont {Cates}},\ }\Doi
  {10.1122/1.551088} {\bibfield  {journal} {\bibinfo  {journal} {J. Rheol.},\
  }\textbf {\bibinfo {volume} {44}},\ \bibinfo {pages} {323} (\bibinfo {year}
  {2000})}\BibitemShut {NoStop}%
\bibitem [{\citenamefont {Jabbari-Farouji}\ \emph {et~al.}(2008)\citenamefont
  {Jabbari-Farouji}, \citenamefont {Atakhorrami}, \citenamefont {Mizuno},
  \citenamefont {Eiser}, \citenamefont {Wegdam}, \citenamefont {MacKintosh},
  \citenamefont {Bonn},\ and\ \citenamefont
  {Schmidt}}]{Jabbari-Farouji_PRE2008}%
  \BibitemOpen
  \bibfield  {author} {\bibinfo {author} {\bibfnamefont {S.}~\bibnamefont
  {Jabbari-Farouji}}, \bibinfo {author} {\bibfnamefont {M.}~\bibnamefont
  {Atakhorrami}}, \bibinfo {author} {\bibfnamefont {D.}~\bibnamefont {Mizuno}},
  \bibinfo {author} {\bibfnamefont {E.}~\bibnamefont {Eiser}}, \bibinfo
  {author} {\bibfnamefont {G.~H.}\ \bibnamefont {Wegdam}}, \bibinfo {author}
  {\bibfnamefont {F.~C.}\ \bibnamefont {MacKintosh}}, \bibinfo {author}
  {\bibfnamefont {D.}~\bibnamefont {Bonn}}, \ and\ \bibinfo {author}
  {\bibfnamefont {C.~F.}\ \bibnamefont {Schmidt}},\ }\Doi
  {10.1103/PhysRevE.78.061402} {\bibfield  {journal} {\bibinfo  {journal}
  {Phys. Rev. E},\ }\textbf {\bibinfo {volume} {78}},\ \bibinfo {pages}
  {061402} (\bibinfo {year} {2008})}\BibitemShut {NoStop}%
\bibitem [{\citenamefont {Willenbacher}(1996)}]{Willenbacher1996}%
  \BibitemOpen
  \bibfield  {author} {\bibinfo {author} {\bibfnamefont {N.}~\bibnamefont
  {Willenbacher}},\ }\href@noop {} {\bibfield  {journal} {\bibinfo  {journal}
  {J. Coll. Int. Sci.},\ }\textbf {\bibinfo {volume} {182}},\ \bibinfo {pages}
  {501} (\bibinfo {year} {1996})}\BibitemShut {NoStop}%
\bibitem [{\citenamefont {Bonn}\ \emph {et~al.}(1998)\citenamefont {Bonn},
  \citenamefont {Tanaka}, \citenamefont {Wegdam}, \citenamefont {Kellay},\ and\
  \citenamefont {Meunier}}]{bonn_epl_1998}%
  \BibitemOpen
  \bibfield  {author} {\bibinfo {author} {\bibfnamefont {D.}~\bibnamefont
  {Bonn}}, \bibinfo {author} {\bibfnamefont {H.}~\bibnamefont {Tanaka}},
  \bibinfo {author} {\bibfnamefont {G.}~\bibnamefont {Wegdam}}, \bibinfo
  {author} {\bibfnamefont {H.}~\bibnamefont {Kellay}}, \ and\ \bibinfo {author}
  {\bibfnamefont {J.}~\bibnamefont {Meunier}},\ }\href@noop {} {\bibfield
  {journal} {\bibinfo  {journal} {Europhys. Lett.},\ }\textbf {\bibinfo
  {volume} {45}},\ \bibinfo {pages} {52} (\bibinfo {year} {1998})}\BibitemShut
  {NoStop}%
\bibitem [{\citenamefont {Joshi}(2007)}]{Joshi_cage_formation_2007}%
  \BibitemOpen
  \bibfield  {author} {\bibinfo {author} {\bibfnamefont {Y.}~\bibnamefont
  {Joshi}},\ }\href@noop {} {\bibfield  {journal} {\bibinfo  {journal} {J.
  Chem. Phys.},\ }\textbf {\bibinfo {volume} {127}},\ \bibinfo {pages} {081102}
  (\bibinfo {year} {2007})}\BibitemShut {NoStop}%
\bibitem [{\citenamefont {Negi}\ and\ \citenamefont
  {Osuji}(2010){\natexlab{a}}}]{Osuji_Negi_arrest_EPL2010}%
  \BibitemOpen
  \bibfield  {author} {\bibinfo {author} {\bibfnamefont {A.}~\bibnamefont
  {Negi}}\ and\ \bibinfo {author} {\bibfnamefont {C.}~\bibnamefont {Osuji}},\
  }\Doi {10.1209/0295-5075/90/28003} {\bibfield  {journal} {\bibinfo  {journal}
  {Europhys. Lett.},\ }\textbf {\bibinfo {volume} {90}},\ \bibinfo {pages}
  {28003} (\bibinfo {year} {2010}{\natexlab{a}})}\BibitemShut {NoStop}%
\bibitem [{\citenamefont {Abou}\ \emph {et~al.}(2001)\citenamefont {Abou},
  \citenamefont {Bonn},\ and\ \citenamefont {Meunier}}]{Bonn_PRE2001}%
  \BibitemOpen
  \bibfield  {author} {\bibinfo {author} {\bibfnamefont {B.}~\bibnamefont
  {Abou}}, \bibinfo {author} {\bibfnamefont {D.}~\bibnamefont {Bonn}}, \ and\
  \bibinfo {author} {\bibfnamefont {J.}~\bibnamefont {Meunier}},\ }\Doi
  {10.1103/PhysRevE.64.021510} {\bibfield  {journal} {\bibinfo  {journal}
  {Phys. Rev. E},\ }\textbf {\bibinfo {volume} {64}},\ \bibinfo {pages}
  {021510} (\bibinfo {year} {2001})}\BibitemShut {NoStop}%
\bibitem [{\citenamefont {Bellour}\ \emph {et~al.}(2003)\citenamefont
  {Bellour}, \citenamefont {Knaebel}, \citenamefont {Harden}, \citenamefont
  {Lequeux},\ and\ \citenamefont {Munch}}]{COO:Bellour_Munch}%
  \BibitemOpen
  \bibfield  {author} {\bibinfo {author} {\bibfnamefont {M.}~\bibnamefont
  {Bellour}}, \bibinfo {author} {\bibfnamefont {A.}~\bibnamefont {Knaebel}},
  \bibinfo {author} {\bibfnamefont {J.~L.}\ \bibnamefont {Harden}}, \bibinfo
  {author} {\bibfnamefont {F.}~\bibnamefont {Lequeux}}, \ and\ \bibinfo
  {author} {\bibfnamefont {J.-P.}\ \bibnamefont {Munch}},\ }\href@noop {}
  {\bibfield  {journal} {\bibinfo  {journal} {Phys. Rev. E.},\ }\textbf
  {\bibinfo {volume} {67}},\ \bibinfo {pages} {031405} (\bibinfo {year}
  {2003})}\BibitemShut {NoStop}%
\bibitem [{\citenamefont {Chow}(1993)}]{Chow1993stress}%
  \BibitemOpen
  \bibfield  {author} {\bibinfo {author} {\bibfnamefont {T.}~\bibnamefont
  {Chow}},\ }\href@noop {} {\bibfield  {journal} {\bibinfo  {journal}
  {Polymer},\ }\textbf {\bibinfo {volume} {34}},\ \bibinfo {pages} {541}
  (\bibinfo {year} {1993})}\BibitemShut {NoStop}%
\bibitem [{\citenamefont {Chen}\ and\ \citenamefont
  {Schweizer}(2008)}]{Schweizer_aging_polymers_PRE2008}%
  \BibitemOpen
  \bibfield  {author} {\bibinfo {author} {\bibfnamefont {K.}~\bibnamefont
  {Chen}}\ and\ \bibinfo {author} {\bibfnamefont {K.~S.}\ \bibnamefont
  {Schweizer}},\ }\Doi {10.1103/PhysRevE.78.031802} {\bibfield  {journal}
  {\bibinfo  {journal} {Phys. Rev. E},\ }\textbf {\bibinfo {volume} {78}},\
  \bibinfo {pages} {031802} (\bibinfo {year} {2008})}\BibitemShut {NoStop}%
\bibitem [{\citenamefont {Miyazaki}\ \emph {et~al.}(2006)\citenamefont
  {Miyazaki}, \citenamefont {Wyss}, \citenamefont {Weitz},\ and\ \citenamefont
  {Reichman}}]{Miyazaki_Wyss2006}%
  \BibitemOpen
  \bibfield  {author} {\bibinfo {author} {\bibfnamefont {K.}~\bibnamefont
  {Miyazaki}}, \bibinfo {author} {\bibfnamefont {H.~M.}\ \bibnamefont {Wyss}},
  \bibinfo {author} {\bibfnamefont {D.~A.}\ \bibnamefont {Weitz}}, \ and\
  \bibinfo {author} {\bibfnamefont {D.~R.}\ \bibnamefont {Reichman}},\
  }\href@noop {} {\bibfield  {journal} {\bibinfo  {journal} {Europhys. Lett.},\
  }\textbf {\bibinfo {volume} {75}},\ \bibinfo {pages} {915} (\bibinfo {year}
  {2006})}\BibitemShut {NoStop}%
\bibitem [{\citenamefont {Derec}\ \emph {et~al.}(2003)\citenamefont {Derec},
  \citenamefont {Ducouret}, \citenamefont {Ajdari},\ and\ \citenamefont
  {Lequeux}}]{Derec_PRE2003}%
  \BibitemOpen
  \bibfield  {author} {\bibinfo {author} {\bibfnamefont {C.}~\bibnamefont
  {Derec}}, \bibinfo {author} {\bibfnamefont {G.}~\bibnamefont {Ducouret}},
  \bibinfo {author} {\bibfnamefont {A.}~\bibnamefont {Ajdari}}, \ and\ \bibinfo
  {author} {\bibfnamefont {F.}~\bibnamefont {Lequeux}},\ }\Doi
  {10.1103/PhysRevE.67.061403} {\bibfield  {journal} {\bibinfo  {journal}
  {Phys. Rev. E},\ }\textbf {\bibinfo {volume} {67}},\ \bibinfo {pages}
  {061403} (\bibinfo {year} {2003})}\BibitemShut {NoStop}%
\bibitem [{\citenamefont {Kaloun}\ \emph {et~al.}(2005)\citenamefont {Kaloun},
  \citenamefont {Skouri}, \citenamefont {Knaebel}, \citenamefont {M\"unch},\
  and\ \citenamefont {H\'ebraud}}]{Hebraud_PRE2005}%
  \BibitemOpen
  \bibfield  {author} {\bibinfo {author} {\bibfnamefont {S.}~\bibnamefont
  {Kaloun}}, \bibinfo {author} {\bibfnamefont {M.}~\bibnamefont {Skouri}},
  \bibinfo {author} {\bibfnamefont {A.}~\bibnamefont {Knaebel}}, \bibinfo
  {author} {\bibfnamefont {J.-P.}\ \bibnamefont {M\"unch}}, \ and\ \bibinfo
  {author} {\bibfnamefont {P.}~\bibnamefont {H\'ebraud}},\ }\Doi
  {10.1103/PhysRevE.72.011401} {\bibfield  {journal} {\bibinfo  {journal}
  {Phys. Rev. E},\ }\textbf {\bibinfo {volume} {72}},\ \bibinfo {pages}
  {011401} (\bibinfo {year} {2005})}\BibitemShut {NoStop}%
\bibitem [{\citenamefont {Viasnoff}\ and\ \citenamefont
  {Lequeux}(2002)}]{Viasnoff_PRL2002}%
  \BibitemOpen
  \bibfield  {author} {\bibinfo {author} {\bibfnamefont {V.}~\bibnamefont
  {Viasnoff}}\ and\ \bibinfo {author} {\bibfnamefont {F.}~\bibnamefont
  {Lequeux}},\ }\href@noop {} {\bibfield  {journal} {\bibinfo  {journal} {Phys.
  Rev. Lett.},\ }\textbf {\bibinfo {volume} {89}},\ \bibinfo {pages} {065701}
  (\bibinfo {year} {2002})}\BibitemShut {NoStop}%
\bibitem [{\citenamefont {Ozon}\ \emph {et~al.}(2003)\citenamefont {Ozon},
  \citenamefont {Narita}, \citenamefont {Knaebel}, \citenamefont {Debr\'egeas},
  \citenamefont {H\'ebraud},\ and\ \citenamefont {Munch}}]{Ozon_PRE2003}%
  \BibitemOpen
  \bibfield  {author} {\bibinfo {author} {\bibfnamefont {F.}~\bibnamefont
  {Ozon}}, \bibinfo {author} {\bibfnamefont {T.}~\bibnamefont {Narita}},
  \bibinfo {author} {\bibfnamefont {A.}~\bibnamefont {Knaebel}}, \bibinfo
  {author} {\bibfnamefont {G.}~\bibnamefont {Debr\'egeas}}, \bibinfo {author}
  {\bibfnamefont {P.}~\bibnamefont {H\'ebraud}}, \ and\ \bibinfo {author}
  {\bibfnamefont {J.-P.}\ \bibnamefont {Munch}},\ }\href@noop {} {\bibfield
  {journal} {\bibinfo  {journal} {Phys. Rev. E},\ }\textbf {\bibinfo {volume}
  {68}},\ \bibinfo {pages} {032401} (\bibinfo {year} {2003})}\BibitemShut
  {NoStop}%
\bibitem [{\citenamefont {Di~Leonardo}\ \emph {et~al.}(2005)\citenamefont
  {Di~Leonardo}, \citenamefont {Ianni},\ and\ \citenamefont
  {Ruocco}}]{Ruocco_PRE2007}%
  \BibitemOpen
  \bibfield  {author} {\bibinfo {author} {\bibfnamefont {R.}~\bibnamefont
  {Di~Leonardo}}, \bibinfo {author} {\bibfnamefont {F.}~\bibnamefont {Ianni}},
  \ and\ \bibinfo {author} {\bibfnamefont {G.}~\bibnamefont {Ruocco}},\ }\Doi
  {10.1103/PhysRevE.71.011505} {\bibfield  {journal} {\bibinfo  {journal}
  {Phys. Rev. E},\ }\textbf {\bibinfo {volume} {71}},\ \bibinfo {pages}
  {011505} (\bibinfo {year} {2005})}\BibitemShut {NoStop}%
\bibitem [{\citenamefont {Negi}\ and\ \citenamefont
  {Osuji}(2010){\natexlab{b}}}]{Osuji_Negi_JRheol2010}%
  \BibitemOpen
  \bibfield  {author} {\bibinfo {author} {\bibfnamefont {A.~S.}\ \bibnamefont
  {Negi}}\ and\ \bibinfo {author} {\bibfnamefont {C.~O.}\ \bibnamefont
  {Osuji}},\ }\Doi {10.1122/1.3460800} {\bibfield  {journal} {\bibinfo
  {journal} {J. Rheol.},\ }\textbf {\bibinfo {volume} {54}},\ \bibinfo {pages}
  {943} (\bibinfo {year} {2010}{\natexlab{b}})}\BibitemShut {NoStop}%
\end{thebibliography}%

\end{document}